\documentstyle[prl,aps,floats,psfig]{revtex}
\begin{document}
\twocolumn[
\hsize\textwidth\columnwidth\hsize\csname@twocolumnfalse\endcsname

\title{Spin-polarized transport in inhomogeneous magnetic 
semiconductors: theory of magnetic/nonmagnetic {\it  p-n} junctions}

\author{Igor \v{Z}uti\'{c}$^1$, Jaroslav Fabian,$^{1,2,}$\cite{address} and 
S. Das Sarma$^{1}$} 
\address{$^1$Department of Physics, University of Maryland at  College
Park, College Park, Maryland 20742-4111, USA\\
$^2$Max-Planck Institute for the Physics of Complex Systems, 
N\"{o}thnitzer Str. 38, D-01187 Dresden, Germany}

\maketitle
 
\begin{abstract}
A theory of spin-polarized transport in inhomogeneous magnetic
semiconductors is developed and applied to magnetic/nonmagnetic
{\it p-n} junctions. Several phenomena with possible spintronic 
applications are predicted, including  spinvoltaic effect, 
spin valve effect, and giant magnetoresistance. It is demonstrated
that only nonequilibrium spin can be injected across the space-charge
region of a {\it p-n} junction, so that there is no spin injection 
(or extraction) at low bias. 
\end{abstract}
\vspace{-0.6cm}
]
Semiconductor spintronics is a developing field 
where active control of spin dynamics is projected to lead to 
device applications with more functionality and better integrability 
with traditional semiconductor technology than metal-based 
spintronics~\cite{dassarma01}. Spin injection into a semiconductor has been 
demonstrated~\cite{fiederling99,hammar99}, but an important question,
what  this injected spin is useful for, remains largely unanswered
(though some semiconductor spintronic device schemes have been 
proposed~\cite{datta90,zutic01}). 
This Letter addresses the problem of spin
and charge bipolar (electron and hole) 
transport in inhomogeneously doped magnetic 
semiconductors (such as BeMnZnSe, CdMnSe, 
or GaMnAs, whose carrier $g$-factors are large due to magnetic doping). 
We show, as an important consequence of our theory, that 
a magnetic/nonmagnetic {\it p-n} junction has significant potential
for device applications, and predict large 
magnetoresistance for such a junction. We also show that 
the magnetic junctions display spinvoltaic and spin valve effects: 
a current, whose direction changes with the direction of applied magnetic field
or injected spin polarization, can flow without any applied bias.
These effects should be useful for sensing magnetic fields, and 
for probing spin polarization and spin relaxation.

To introduce the equations for spin and charge 
bipolar transport in inhomogeneously doped magnetic semiconductors,
consider a semiconductor doped with $N_a({\bf r})$ acceptors and 
$N_d({\bf r})$ donors, and with 
magnetic impurities, whose density varies in space 
and whose presence leads to large $g$-factors for electrons and holes,
$g_n({\bf r})$ and $g_p({\bf r})$. In a homogeneous magnetic field $B$ 
the carrier energies are Zeeman-split:
spin up ($\lambda=1$ or $\uparrow$) and spin down
($\lambda=-1$ or $\downarrow$) electrons have their energy shifted by
$-\lambda q\zeta_n({\bf r})=-\lambda g_n({\bf r})\mu_B B/2$, where $\mu_B$ is
Bohr magneton and $q$ is the proton charge; the energy of holes 
changes by
$\lambda q\zeta_p({\bf r})=\lambda g_p({\bf r})\mu_B B/2$.
Carrier charge current densities, resulting from 
the electric field ${\bf E}=-\nabla\phi$, 
nonuniform magnetic ``potentials'' $\zeta$, electron (hole) 
densities $n$ ($p$), are 
\begin{eqnarray}\label{eq:Jn} 
{\bf J}_{n\lambda}&=
&q\mu_{n\lambda}n_{\lambda}{\bf E}+qD_{n\lambda}\nabla n_\lambda
-q\lambda\mu_{n\lambda}n_{\lambda}\nabla \zeta_n,\\ \label{eq:Jp}
{\bf J}_{p\lambda}&=&
q\mu_{p\lambda} p_{\lambda} {\bf E}-qD_{p\lambda} \nabla p_\lambda
-q\lambda\mu_{p\lambda}p_{\lambda}\nabla \zeta_p,
\end{eqnarray}
where $\mu$ and $D$ stand for mobility and diffusivity,
and  the ``magnetic'' drift $q\nabla\zeta$ forces carriers with opposite 
spins to go in opposite directions. The above equations follow from the 
requirement that carrier densities in an inhomogeneous environment 
have the quasiequilibrium form, for example, 
$n_{\lambda}=n_{\lambda}(q\phi+\lambda q\zeta_n +\tilde{\mu}_{\lambda})$, 
where $n_{\lambda}(\tilde{\mu}_{\lambda})$ is the number of electrons per 
spin $\lambda$ at (nonequilibrium) chemical potential $\tilde{\mu}_{\lambda}$,
and the expression for the current $J_{n\lambda}=n_\lambda
\mu_{\lambda}\nabla\tilde{\mu}_{\lambda}$~\cite{tiwari92}. 
From  Eqs.~\ref{eq:Jn} and \ref{eq:Jp}, the carrier charge and 
spin current densities
$J=J_{\uparrow}+J_{\downarrow}$ and $J_s=J_{\uparrow}-J_{\downarrow}$ are:
\begin{eqnarray}\label{eq:Jnt}
{\bf J}_n&=
&\sigma_n E- \sigma_{sn} \nabla\zeta_n + qD_n\nabla n+qD_{sn}\nabla s_n,\\ 
\label{eq:Jpt}
{\bf J}_p&=
&\sigma_p {\bf E}- \sigma_{sp} \nabla \zeta_p-qD_p\nabla p-qD_{sp}\nabla s_p,
\end{eqnarray} 
and
\begin{eqnarray}\label{eq:Jnst}
{\bf J}_{sn}&=
&\sigma_{sn} {\bf E}- \sigma_n \nabla  \zeta_n+ qD_{sn} 
\nabla n+qD_n \nabla s_n,\\ \label{eq:Jpst}
{\bf J}_{sp}&=
&\sigma_{sp} {\bf E}- \sigma_p \nabla \zeta_p- qD_{sp} 
\nabla p-qD_p \nabla s_p.
\end{eqnarray}
Here $n=n_{\uparrow}+n_{\downarrow}$ and $s_n=n_\uparrow-n_\downarrow$, 
and we introduced electron charge and spin conductivities 
$\sigma_n=q(\mu_nn+\mu_{sn}s_n)$  
and $\sigma_{sn}=q(\mu_{sn}n+\mu_ns_n)$, 
where $\mu_n=(\mu_{n\uparrow}+ 
\mu_{n\downarrow})/2$ and $\mu_{sn}=(\mu_{n\uparrow}-\mu_{n\downarrow})/2$, 
and similarly
for diffusivities. Analogous notation is used for holes.
Equations \ref{eq:Jnt}-\ref{eq:Jpst}, which are a generalization of
the Johnson-Silsbee magnetotransport equations~\cite{johnson87}, 
reflect the spin-charge coupling in
bipolar transport in inhomogeneous magnetic semiconductors: 
a spatial variation in spin density, as well as in  
$q\nabla\zeta$, can cause charge currents, and spin currents can flow as a 
result of a spatial variation of carrier densities and $\phi$. 

\begin{figure}
\centerline{\psfig{file=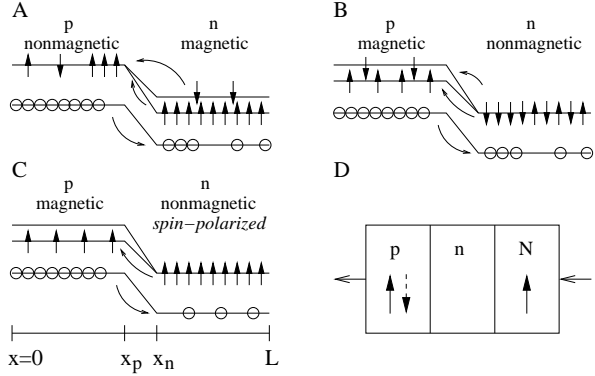,width=1\linewidth,angle=-90}}  
\vspace{-0.3cm}
\caption{Band-energy schemes for magnetic/nonmagnetic
{\it p-n} junctions with magnetically active electrons (arrows). Holes
(circles) are unpolarized. Electrons from the magnetically active 
$n$-region (A) (discernible by the conduction band split) can be 
injected into the nonmagnetic $p$-region only at large bias; similarly
for spin extraction from a magnetic $p$-region (B), where electrons
are minority carriers. If spin of the majority
electrons is out of equilibrium (they are spin polarized),
and the $p$-region is magnetic (C),
a giant magnetoresistance and spinvoltaic effects arise. 
The former can be observed in a scheme (D), where electron spin is injected
from a magnetic heterostructure $N$ into the nonmagnetic $n$-region, which
forms a {\it p-n} junction with a magnetic $p$-region.
}
\label{fig:1}
\vspace{-0.3cm}
\end{figure}

Generation and recombination of electrons and holes are assumed to be
mostly due to band-to-band processes. Furthermore, electrons (holes) 
with a given spin are assumed to recombine with holes (electrons) of 
either spin.  The stationary continuity 
equations for electrons and holes read
\begin{eqnarray}\label{eq:wn}
\nabla\cdot \frac{{\bf J}_{n\lambda}}{q}&=&+
w_{n\lambda}(n_{\lambda}p-n_{\lambda0}p_0)+
\frac{n_{\lambda}-n_{-{\lambda}}-\lambda \tilde {s}_{n}}{2T_{1n}},\\
\label{eq:wp}
\nabla\cdot \frac{{\bf J}_{p\lambda}}{q}&=&-
w_{p\lambda}(p_{\lambda}n-p_{\lambda 0}n_0)-
\frac{p_{\lambda}-p_{-{\lambda}}-\lambda \tilde {s}_{p}}{2T_{1p}}.
\end{eqnarray}
Here $n_0$ and $p_0$ are the equilibrium carrier densities, $w$ is
the (generally spin dependent) band-to-band recombination rate, 
and $T_1$ is the spin relaxation time. Spin relaxation equilibrates electron
spins while preserving {\it nonequilibrium} electron  density, 
so, for a nondegenerate semiconductor, $\tilde{s}_{n}=\alpha_{n0}n$, 
where $\alpha_{n0}=s_{n0}/n_0=\tanh(\zeta_n/V_T)$ is
the equilibrium electron spin polarization ($V_T=k_B T/q$, with $k_B$ being 
the Boltzmann constant and $T$ temperature).  
In general, $\tilde{s}_{n}=n(\tilde{\mu}+q\zeta_n)-n(\tilde{\mu}-q\zeta_n)$,
where $\tilde{\mu}$ is obtained from the 
known nonequilibrium electron density $n=n(\tilde{\mu}+q\zeta_n)+
n(\tilde{\mu}-q\zeta_n)$; similarly for holes. To include transient effects, terms  
$-\partial n_{\lambda}/\partial t$ and $\partial p_{\lambda}/\partial t$ 
need to be added to the left sides of
Eqs.~\ref{eq:wn} and \ref{eq:wp}.


\begin{figure}
\centerline{\psfig{file=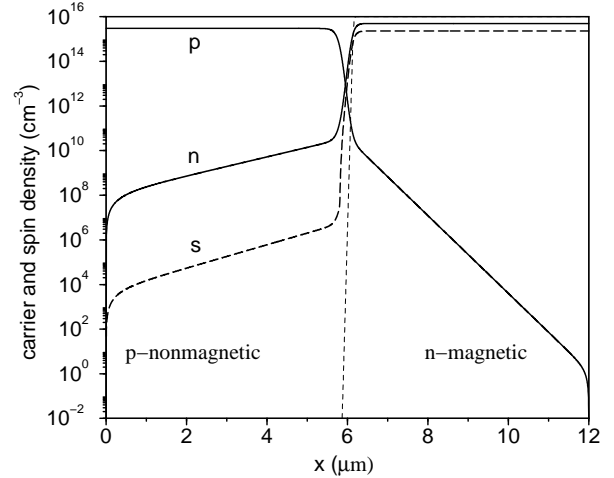,width=1\linewidth,angle=-90}}  
\vspace{0.2cm}
\caption{Calculated electron ($n$), hole ($p$), and spin ($s$) density profiles
for a magnetic {\it p-n} junction with $q\zeta=0.5 k_B T$ in the (magnetic) 
$n$-region, under forward applied bias of $V=0.8$ Volt. The thin dashed line 
indicates the doping
profile $N_d-N_a$ (not to scale).}
\label{fig:2}
\vspace{-0.3cm}
\end{figure}

Equations \ref{eq:Jn} and \ref{eq:Jp} 
(or, \ref{eq:Jnt}-\ref{eq:Jpst}), \ref{eq:wn} and
\ref{eq:wp}, and Poisson's equation 
$\nabla\cdot \epsilon {\bf E}=\rho$, 
where $\rho=q(p-n+N_d-N_a)$ and $\epsilon$ 
is the dielectric constant of the semiconductor, 
determine the distributions of charge and spin
in a magnetic semiconductor under applied bias V. 
For ferromagnetic semiconductors $\zeta$ needs to be evaluated 
self-consistently 
(similarly to $\phi$). Indeed, the magnetization depends on the carrier 
density (as the mobile carriers mediate the magnetic interactions), and
the carrier density depends on magnetization, as magnetization determines
$\zeta$. 

We have made a number of simplifying assumptions in our theory which are
not strictly valid in real magnetic semiconductors such as GaMnAs.
For example, electron and hole band states are treated as simple spin 
doublets, although the band structure is usually more complicated
in the presence of spin-orbit coupling. 
We also do not take into account the fact that magnetic impurities
change the gap and can offset the bands, even at B=0. Finally, in some magnetic semiconductors
(even nonferromagnetic) carrier densities may strongly affect the
 response
of the magnetic ions to magnetic field, in which case a selfconsistent 
calculation
is needed. In principle,  our formalism can be tailored to 
include
these more complicated
effects, but this can only be done on a case-by-case basis (if the 
effects
are well understood, which is currently not the case). Our approach here
is to consider a simple but reasonable model to illustrate new phenomena that 
can occur in inhomogeneous magnetic semiconductors. 

We now apply the above theory to the problem of a magnetic/nonmagnetic 
{\it p-n} junction, with only electrons being magnetically active 
($\zeta_p=0$)~\cite{holes}.
Two cases are considered (Fig.~\ref{fig:1}): 
(i) the magnetic $n$-side to study spin 
injection, and (ii) the magnetic $p$-side, to study spin extraction
and magnetoresistance phenomena. We assume perfect ohmic contacts 
(both carriers and spins at equilibrium) as boundary conditions, 
except when we study the dependence of the current on nonequilibrium spin 
(in which case we keep the 
ohmic contacts for carrier densities only, not for the spin). We only
deal with nondegenerate semiconductors, as these are analytically
tractable, and yield the largest magnetoresistance. 

\begin{figure}
\centerline{\psfig{file=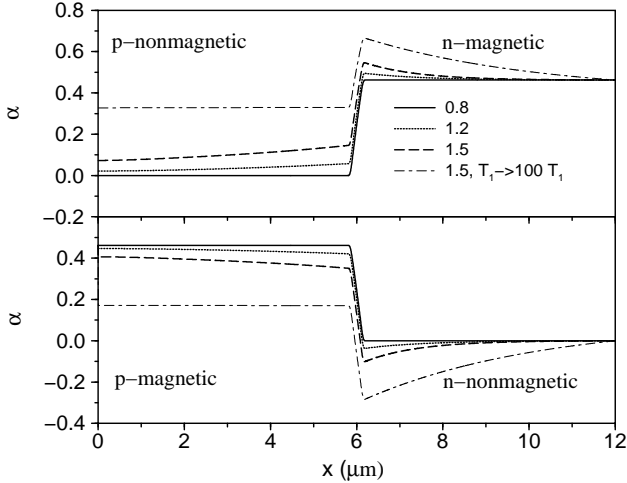,width=1\linewidth,angle=-90}}      
\vspace{0.2cm}
\caption{Calculated spin polarization profiles for different forward bias and
spin relaxation rate. Spin injection (top) from the magnetic $n$-side into
the nonmagnetic $p$-side occurs only at large bias (indicated by the numbers
in Volts). The largest injection in the graph is for $V=1.5$ Volt, with
spin relaxation time of $100 T_1$. Similar behavior is observed for spin
extraction from the nonmagnetic $n$-region into the magnetic $p$-region
(bottom). The magnetic splitting is $q\zeta=0.5 k_B T$. }
\label{fig:3}
\vspace{-0.3cm}
\end{figure}

To be specific, we consider a $L=12$ $\mu$m long GaAs {\it p-n} junction at 
room temperature,  
doped with $N_a=3\times 10^{15}$ cm$^{-3}$ acceptors to the left, 
$N_d=5\times 10^{15}$ cm$^{-3}$ donors to the right~\cite{zutic01}, 
and with magnetic impurities
inducing electronic $g$-factor which follows the shape of $N_a$ ($N_d$) 
for the magnetic $p$ ($n$) region. Holes are unpolarized, so we
omit the label $n$ when dealing with spin-related quantities, and reserve
that symbol for the $n$-region-related variables. Figure~\ref{fig:1}
depicts our geometry and the distribution of electrons and holes
for various cases considered below.
The materials parameters~\cite{parameters} are the electron and hole 
diffusivities, $D_n=10 D_p=103.6$ cm$^2$$\cdot$s$^{-1}$,
and mobilities $\mu_n=10\mu_p= 4000$ cm$^2\cdot$V$^{-1}\cdot$s$^{-1}$. 
The intrinsic (nonmagnetic) carrier density is $n_i=1.8\times 10^{6}$ cm$^{-3}$, 
the dielectric constant $\epsilon=13.1$ of the vacuum permittivity; the
calculated built-in-voltage is $1.1$ V at $B=0$.
Recombination rate $w=(1/3)\times 10^{-5}$ cm$^3\cdot$s$^{-1}$, and
the spin relaxation time $T_1=0.2$ ns. The minority~\cite{minority}
diffusion lengths 
are~\cite{zutic01} $L_n=1$ $\mu$m, $L_p=0.25$ $\mu$m, and the electron 
spin diffusion length in the $n$ ($p$) region is $L_{sn}=1.4$ $\mu$m 
($L_{sp}$=0.8 $\mu$m). Figure~\ref{fig:2} illustrates the doping profile 
and a typical profile of carrier and spin densities.

We first ask the important question whether spin can be injected and extracted 
into/from
the nonmagnetic region. Figure~\ref{fig:3} shows the results of our 
numerical calculations, where we plot spin polarization $\alpha=s/n$ (note
that we are interested here in spin polarization of density, 
not current~\cite{zutic01}).
At small bias (below the built-in-value), which is also 
the small injection limit, there is no significant spin injection or extraction.
As the bias increases, the injection and extraction become large
and intensify with increasing $T_1$. 
The reason why there is no spin injection (and, similarly, 
extraction) at small bias is that although there are exponentially
more, say, spin up than spin down electrons in the magnetic side, the barrier
for crossing the space-charge region is exponentially larger for spin up
than for spin down electrons (see Fig.~\ref{fig:1}). 
Those two exponential effects cancel each other, so there is no net spin 
current flowing through the space-charge region.
The numerical aproach is indispensible for obtaining the high-injection
results, where we observe a large nonequilibrium spin polarization around
the space-charge region, resulting in spin injection and extraction. 

\begin{figure}
\centerline{\psfig{file=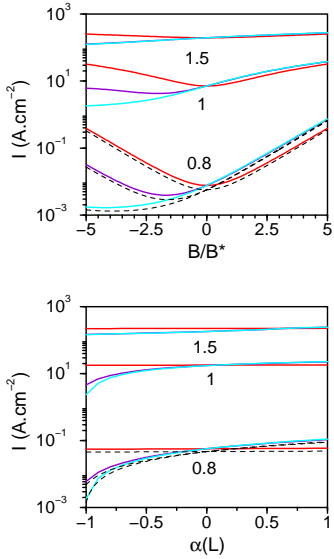,width=1.0\linewidth,angle=-90}} 
\vspace{0.5truecm}
\caption{Calculated $I$-$B$ (top) and $I$-$\alpha$ (bottom) 
characteristics for the
magnetic {\it p-n} junction with the magnetic $p$-region,
under different bias (indicated in Volts). We use $B^*=2 k_B T/ g \mu_B$ 
($\approx 900/g$ Tesla at 
room temperature) as the scale for magnetic field ($B/B^*=q\zeta/V_T$).
In the top graph the red line is for $\alpha(L)=0$, and the violet (cyan)  
line for $\alpha=1$ and spin relaxation
time of 100 ($10^4$) $T_1$. The bottom graph is for $B/B^*=3$, and the
red, violet, and cyan lines are for the spin relaxation times of
$T_1$, $100 T_1$, and $10^4 T_1$. The dashed lines
are the analytical calculations. In both graphs, at $V=1.5$ Volt,
lines for $100 T_1$ and $10^4 T_1$ almost coincide.
}
\label{fig:4}
\vspace{-0.3cm}
\end{figure}

The current through a magnetic/nonmagnetic {\it p-n} junction depends on
magnetic field. This dependence has two sources. First, a magnetic field
makes the band gap spin dependent, leading to the increased density of
the minority carriers. For example, the equilibrium density of electrons
in a magnetic $p$-region (Fig.~\ref{fig:1}C) is 
$n_0=n_i^2\cosh(\zeta/V_T)/N_a$ 
(in equilibrium, $n_0p_0=n_i^2\cosh(\zeta/V_T)$ from Boltzmann statistics).
The current through a {\it p-n} junction is proportional to the density
of the minority (not majority) carriers, and so it will grow exponentially
with $B$ for $\zeta \agt V_T$. This does not happen  for unipolar
transport (a homeogenous sample), where the current depends on the density of 
the majority carriers, which is independent of $B$, as it is fixed by the 
number of ionized impurities. The exponential magnetoresistance for the 
system in Fig.~\ref{fig:1}B is shown in Fig.~\ref{fig:4}. 
The effect diminishes with increasing bias. The second way a current can 
depend on $B$ is if a nonequilibrium spin with polarization $\alpha(L)$ 
is introduced (optically or by spin injection) into the sample 
(Fig.~\ref{fig:1}C and D). The exponential magnetoresistance becomes giant
(meaning, the resistance changes when the orientation of the magnetic
moment in the magnetic region with respect to the orientation of
the injected spin in the nonmagnetic region, changes)
as seen in Fig.~\ref{fig:4}. The current is exponentially sensitive to both
$B$ and $\alpha(L)$, being large (small) if they are of the same (opposite) 
sign.

To explain analytically our numerical results, we generalize the 
well-known Shockley approximation~\cite{tiwari92}
for the {\it p-n} junction to include nonequilibrium 
spin, which is given, in the quasi-neutral regions, by the spin-diffusion 
equation
\begin{eqnarray}
D_n\nabla^2 s=w(sp-s_0p_0)+(s-\tilde{s})/T_1,
\end{eqnarray} 
where $\tilde{s}=\alpha_0 n$, of which $n$ is the solution of the
corresponding carrier diffusion equation.
We consider the magnetic
$p$-region, and impose a nonequilibrium spin polarization $\alpha(L)$
at the right boundary (see Fig.~\ref{fig:1}C, where also the  
boundary points used below are indicated). 
We {\it postulate} as the second boundary condition for spin, 
in the spirit of the quasiequilibrium
approximation~\cite{tiwari92}, that spin 
current vanishes at $x_n$, the $n$-side boundary of the space-charge region.
The spin polarization in the $n$-region is 
\begin{eqnarray}
\alpha(x)=\alpha(L) \left [\cosh(\eta)-\tanh(\eta_n)\sinh(\eta)\right ],
\end{eqnarray}
where $\eta=(L-x)/L_{sn}$ and $\eta_n=(L-x_n)/L_{sn}$.
At $x=x_n$,  $\alpha_n\equiv\alpha(x_n)=\alpha(L)/\cosh\left 
[(L-x_n)/L_{sn}\right ]$; 
it vanishes if $\alpha(L)=0$, in line with our result of no spin extraction 
at low bias.  In the $p$-region $\alpha(0)=\alpha_0$ (ohmic boundary
condition for spin) and the second boundary 
condition, that at $x=x_p$, the $p$-side
boundary of the space-charge region, which can be obtained from the usual 
quasiequilibrium condition of constant chemical potentials (applied
to nondegenerate statistics) from $x_p$ to $x_n$, is  
$\alpha_p=(\alpha_0+\alpha_n)/(1+\alpha_0\alpha_n)$. The nonequilibrium minority
carrier density at $x=x_p$ is  $n(x_p)=n_0 (1+\alpha_0\alpha_n)\exp(V/V_T)$.
With these boundary conditions the carrier and spin profiles in the
magnetic $p$-region are
\begin{eqnarray}
\frac{n(x)}{n_0}&=& 1 + \frac{\sinh(x/L_n)}{\sinh(x_p/L_n)}
\left [e^{V/V_T}
(1+\alpha_n\alpha_0)-1\right ],\\
\frac{s(x)}{s_0}&=&\frac{n(x)}{n_0}+\frac{\sinh\left (x/L_{sp}\right )}
{\sinh\left (x_p/L_{sp}\right )}
\frac{\alpha_n}{\alpha_0}(1-\alpha_0^2)e^{V/V_T}.
\end{eqnarray}

Having the profiles, we calculate the current in the magnetic 
{\it p-n} junction. We distinguish equilibrium-spin electron $J_n$ and
hole $J_p$ currents,
and nonequilibrium-spin-induced current $J_n'$,
so that the total charge current is $J=J_n+J_p+J_n'$. The individual
contributions are 
\begin{eqnarray} \label{eq:J1}
J_{n}&=&q\frac{D_n}{L_n} \frac{n_i^2}{N_a}\cosh\left (\frac{\zeta}{V_T}\right )
 \coth\left (\frac{x_p}{L_n}\right ) \left (e^{V/V_T}-1\right ),\\ \label{eq:J2}
J_{p}&=&q\frac{D_p}{L_p} \frac{n_i^2}{N_d}\coth\left (\frac{L-x_n}{L_p}\right ) 
\left (e^{V/V_T}-1\right ),\\ \label{eq:J3}
J_{n}'&=&q \frac{D_n}{L_n} \frac{n_i^2}{N_a} \coth\left (\frac{x_p}{L_n}\right ) 
\sinh\left (\frac{\zeta}{V_T}\right )\alpha_n
e^{V/V_T}.
\end{eqnarray}
Figure~\ref{fig:4} shows how well this analytical model compares
with our numerical calculation. The small discrepancy
is caused by the neglect of the recombination processes inside the 
space-charge region; the model breaks down at large bias, where 
the Shockley approximation becomes invalid~\cite{tiwari92}. 
We note that if the spin is injected by
the bias contact at $x=L$, one needs to consider a model in which $\alpha(L)$
depends on $J$ (since it then obviously vanishes at $J=0$), and solve 
Eqs.~\ref{eq:J1}-\ref{eq:J3} for $J$.  Otherwise the above equations describe
either the case of an independent spin injection, or the region of $J$ where
$\alpha(L)$ is independent of  $J$.

The large exponential magnetoresistance effect, $J\sim \exp(|\zeta|/V_T)$ at 
large 
$|\zeta|$ (or $B$), 
comes from the increase of the minority electron population with $|\zeta|$.  
However, once a nonequilibrium spin population (finite $\alpha_n$) is 
maintained with its sign fixed in the space-charge region, 
a giant magnetoresistance (GMR) should be observed,
$J\sim \exp(\zeta/V_T)$ at large $|\zeta|$ and $\alpha_n\rightarrow\pm 1$.
The GMR coefficient is then $\exp(2\zeta/V_T)$. In addition
to sensing $B$, these effects can be used for an all-electrical
probing of the injected spin polarization $\alpha(L)$ 
and spin diffusion 
length $L_{sn}$, as they both determine $\alpha_n$ (a similar device
scheme is proposed in Fig.~\ref{fig:1} D). 
The key to employing the
large magnetoresistance effects in spintronics is the development of
 materials with large $g$-factors. Even  for  
$g\approx 100$, the GMR coefficient at $T=100$ K and $B=1$T is about 
$\exp{(0.65)}$, which is close to 200\%.
A magnetic impurity that would give $g\approx 500$ would yield 2,500 \%,
(300\% at room temperature). The device potential of magnetic {\it p-n}
junctions is enormous, but to take full advantage we need materials
with large $g$-factors at room temperature~\cite{dietl94}.

Equations \ref{eq:J1}-\ref{eq:J3} reveal another interesting phenomenon,
a spinvoltaic effect. While both
$J_n$ and $J_p$ vanish for $V=0$, $J_n'$ stays finite, if $\alpha_n\ne 0$.
A current flows at zero bias! 
This is an analogue of the photovoltaic effect, where a reverse current can 
flow
by introducing photogenerated carriers into the space charge region.
In the spinvoltaic effect both reverse and forward currents can flow, depending
on the relative orientation of $B$ and $\alpha_n$, 
so a magnetic {\it p-n} junction can act as a spin valve. 
The physics of the spinvoltaic effect is that nonequilibrium spin 
in the space-charge region disturbs the balance between the generation and 
recombination currents (Fig.~\ref{fig:1}C). 
If $\zeta > 0$, and more spin up electrons are
present at $x_n$ ($\alpha_n > 0$), the barrier for them to cross the region
is smaller than the barrier for the spin down electrons, so more electrons
flow from $n$ to $p$ than from $p$ to $n$, and positive charge current 
results.
If there are more spin down electrons at $x_n$ ($\alpha_n < 0$), the current
is reversed. We note that, as at $V=0$ no spin can be injected by the ohmic
contact into the $n$ region, the nonequilibrium spin population has to be 
injected either optically (at a distance larger than $L_n$ but smaller than
$L_{sn}$ from the space-charge region), or electronically transverse to the
flow of $J$ (and also within $L_{sn}$ of the space-charge region).

In summary, we have explored bipolar magnetotransport in magnetic/nonmagnetic
{\it p-n} junctions, which are of great potential for emerging semiconductor 
spintronics. We have used a simple model to describe the new physics that results 
when the magnetotransport is dominated by minority, rather than majority 
charge carriers. 
In particular, we find that spin injection is suppressed at low biases,
because the spin-dependent barrier to transport in the depletion layer negates
the effect of spin polarization of the carriers. Furthermore, we predict 
exponential magnetoresistance resulting from the magnetic field dependence of the 
population of the minority carriers,  and giant magnetoresistance,
appearing if the polarization of the majority carriers is disturbed
externally (by a third-terminal spin injection, for example).
We also find that a nonequilibrium spin-polarized
population can lead to a charge current--a spinvoltaic effect. 
Finally, we have shown that at small biases the effects of interest can
be described by a modified Shockley model.

This work was supported by DARPA and the US ONR.


\begin{references}
\bibitem{holes} We neglect hole spins in our model for the sake
of clarity so that we
illustrate the results on one carrier type. It is straightforward to
include hole spin transport (especially in our analytical formulas).
\bibitem{parameters} Chosen to have typical values for GaAs with
no special significance about the numbers we use.
\end{references}
\end{document}